\documentclass[aps,pra,superscriptaddress,twocolumn,longbibliography]{revtex4-1}
\usepackage{amsmath}
\usepackage[english]{babel}
\usepackage{bbm}
\usepackage[normalem]{ulem}
\usepackage{times}
\usepackage[normalem]{ulem}
\usepackage{dsfont}
\usepackage{txfonts}
\usepackage{graphicx}
\usepackage{dcolumn}
\usepackage{bm}
\usepackage{tabularx}
\usepackage{enumerate}
\usepackage[colorlinks=true, citecolor=blue,urlcolor=blue, linkcolor=blue,citebordercolor={1 0 0},linkbordercolor={0 0 1}]{hyperref}
\usepackage{amsfonts,amssymb}
\usepackage{mathrsfs}
\usepackage{subfigure}
\usepackage{pifont}
\usepackage{cleveref}
\usepackage{color}
\usepackage{thmtools}
\usepackage{verbatim}
\usepackage[ruled,vlined]{algorithm2e}

\newcommand{\bra}[1]{\langle#1|}

\newcommand{\per}{\mathcal{P}}

\newcommand{\ket}[1]{|#1\rangle}

\newcommand{\pcut}{p_{\text{cut}}}
\newcommand{\pmax}{p_{\text{max}}}
\newcommand{\pavg}{p_{\text{avg}}}

\newcommand{\aabb}{\alpha\alpha\beta\beta}
\newcommand{\abab}{\alpha\beta\alpha\beta}
\newcommand{\beh}{\mathrm{BeH_2}}

\newcommand{\equref}[1]{Eq.~(\ref{#1})}

\newcommand{\utchem}{Department  of  Chemistry,  University  of  Toronto,  Toronto,  Ontario  M5G 1Z8,  Canada}
\newcommand{\utcomp}{Department  of  Computer Science,  University  of  Toronto,  Toronto,  Ontario  M5S 2E4,  Canada}
\newcommand{\vectorinst}{Vector  Institute  for  Artificial  Intelligence,  Toronto,  Ontario  M5S  1M1,  Canada}
\newcommand{\cifar}{Canadian  Institute  for  Advanced  Research,  Toronto,  Ontario  M5G  1Z8,  Canada}
\newcommand{\sustech}{Department of Physics, Southern University of Science and Technology, Shenzhen, 518055, China}

\begin{document}

\title[]{Mutual information-assisted Adaptive Variational Quantum Eigensolver}

\author{Zi-Jian Zhang}
\address{\sustech}
\address{\utcomp}
\address{\utchem}

\author{Thi Ha Kyaw}
\address{\utcomp}
\address{\utchem}

\author{Jakob S. Kottmann}
\address{\utcomp}
\address{\utchem}

\author{Matthias Degroote}
\address{\utcomp}
\address{\utchem}

\author{Al\'{a}n Aspuru-Guzik}
\address{\utcomp}
\address{\utchem}
\address{\vectorinst}
\address{\cifar}

\date{\today}

\begin{abstract}    
Adaptive construction of ansatz circuits offers a promising route towards applicable variational quantum eigensolvers on near-term quantum hardware. 
Those algorithms aim to build up optimal circuits for a certain problem and ansatz circuits are adaptively constructed by selecting and adding entanglers from a predefined pool. 
In this work, we propose a way to construct entangler pools with reduced size by leveraging classical algorithms.
Our method uses mutual information between the qubits in classically approximated ground state to rank and screen the entanglers. 
The density matrix renormalization group method is employed for classical precomputation in this work.
We corroborate our method numerically on small molecules.
Our numerical experiments show that a reduced entangler pool with a small portion of the original entangler pool can achieve same numerical accuracy.
We believe that our method paves a new way for adaptive construction of ansatz circuits for variational quantum algorithms.

\end{abstract}

\maketitle

\section{Introduction}
Quantum computers promise speed-up for solving certain computational problems~\cite{nielsen2002quantum,shor1994algorithms,harrow2009quantum} over their classical counterparts.
Despite recent progresses~\cite{arute2019quantum,wright2019benchmarking,PhysRevLett.122.110501} in quantum computing hardware, we remain in the noisy intermediate-scale quantum (NISQ) era~\cite{preskill2018quantum} where the number of qubits and the depth of quantum circuit are limited to a few tens of qubits and gates.
These limitations make it pressing to find problems and algorithms suitable for the NISQ devices. 
It is believed that the electronic structure problem in quantum chemistry~\cite{atkins2011molecular,helgaker2014molecular} is a good problem for NISQ devices~\cite{Reiher7555}.
There exists a number of quantum algorithms~\cite{cao2019quantum,mcardle2018quantum,peruzzo2014variational,McClean_2016,PhysRevLett.83.5162,Aspuru-Guzik1704} for solving the electronic structure problem; and the variational quantum eigensolver (VQE)~\cite{peruzzo2014variational,McClean_2016} is one of them which requires a feasible number of quantum gates for near-term devices and has noise resilient feature.

Currently one major research area of the VQE algorithms is the ansatz design. 
To maximally utilize available quantum hardware resources, the hardware-efficient ansatz~\cite{kandala2017hardware} was introduced. 
In this ansatz, single-qubit unitary gates and a fixed entangling unitary that is easy to implement on the corresponding hardware are placed alternately.
Another important family of ansatz in this field is unitary coupled cluster (UCC)~\cite{Romero_2018}, which was inspired by the coupled cluster method in classical computational chemistry.
However, the hardware-efficient ansatz and the UCC ansatz often employ too many quantum gates and hence the circuit depth can become too large even for a small system size.
Hence, they become problematic in experiments, since the NISQ devices are noisy and only have short coherence time.
Recently, a number of adaptive ansatz construction methods, such as the qubit coupled cluster (QCC) methods~\cite{ryabinkin2018qubit,ryabinkin2020iterative} and the ADAPT-VQE methods~\cite{grimsley2019adaptive,tang2019qubit}, have been proposed to overcome this challenge. 
In these methods, the ansatz is iteratively constructed based on the previous results of circuit runs.
The adaptiveness of these methods lets them place quantum gates in the most appropriate places and thus reduces the number of gates involved. 
Nonetheless, in some cases, the adaptive construction needs many circuit runs, thereby negating its computational efficiency.

On the other hand, mutual information (MI) between spin-orbitals has long been used to improve classical computational chemistry algorithms~\cite{rissler2006measuring,legeza2003optimizing,stein2019autocas}. 
In the density matrix renormalization group (DMRG) algorithm for quantum chemistry~\cite{chan2011density}, the molecular orbitals are mapped to an artificial one-dimensional spin chain. 
This mapping is not unique and the performance of the algorithm depends on the choice of the mapping. 
The MI between the orbitals can be used to iteratively improve the mapping.
It is well-known that even half converged DMRG calculation could provide a useful MI estimation, which in turn helps to converge the DMRG with less computational resources~\cite{legeza2003optimizing}.

In this work, we propose a method to reduce the size of the entangler pools and thereby the number of circuit runs needed in the adaptive ansatz construction methods. 
Our method makes use of approximated MI from classical methods such as DMRG. 
We carry out numerical experiments of our method on the Hamiltonians of the $\mathrm{H_2}$, $\mathrm{LiH}$, $\mathrm{H_2O}$ and $\mathrm{BeH_2}$ molecules and show how the QCC and similar methods can be accelerated.
Our results show that the proposed MI-assisted adaptive VQE can significantly reduces the required number of the circuit runs, thereby achieving a significant speed-up over the existing methods.

\section{Methods}
\label{sec:methods}
\subsection{Adaptive ansatz construction in VQE}
\label{subsec:adaptive_VQE}
Variational quantum eigensolver is a class of quantum algorithms aiming to solve the ground state energy of general qubit Hamiltonians which can be decomposed as
$
	\hat{H} = \sum_i a_i \hat{P}_i,
$
where $\hat{P}_i$ are tensor products of Pauli operators, which we will call Pauli words onwards.
During a VQE run, parameterized trial wavefunctions $\ket{\Psi(\vec{\varphi})}=\hat{U}(\vec{\varphi})\ket{\Psi_0}$ are prepared on a quantum computer, where the reference state $\ket{\Psi_0}=\hat{U}_0\ket{0}$ is usually chosen to be the Hartree-Fock state. 
Typically, the goal of VQE is to obtain the ground state energy of the given Hamiltonian. To achieve this, one needs to update $\vec{\varphi}$ using a classical computer and try to find the minimum of the energy estimation, which is expectation value of the Hamiltonian:
\begin{equation}
    E_{\min} = \min_{\vec{\varphi}}\bra{\Psi(\vec{\varphi})}\hat H\ket{\Psi(\vec{\varphi})}.
\end{equation}

Within adaptive ansatz construction methods, the trial wavefunction or ansatz is usually written as
\begin{equation}
\ket{\Psi(\vec{\varphi})}=\prod_{i=1}^{N_{\mathrm{ent}}} \hat{U}_i(\vec{\varphi_i}) \ket{\Psi_0},
\end{equation}
where the entanglers $\hat{U_i}$ are chosen from an entangler pool $E=\{\hat{E}_1, \hat{E}_2,\cdots,\hat{E}_{m}\}$.
Here, we use $\hat{U}$ to represent the entanglers in the circuit and $\hat{E}$ to represent the entanglers in the pool. 
As the name ``adaptive ansatz construction" suggests, it relies on flexible ansatz, constructed stepwise. 
Typically at any $n^{\mathrm{th}}$ step of the ansatz construction, 
entanglers in the pool are scored based on some measure and an entangler $\hat{E}_i\in E$ with the highest score is selected to be added to the ansatz as $\hat{U}_n$. 
A typical score is the gradient of the energy with respect to the parameters in the entangler. This score is adopted in QCC and ADAPT-VQE. The score can be also defined as the amount of energy that can be reduced by adding the entangler $\hat{E}_i$. In this work we adopt the second score.
We formally define the adaptive ansatz construction process used in this work below.
\begin{enumerate}
    \item Define the initial state $\ket{\Psi_0}$ and the entangler pool $E$.
    \item For the $n^{\textrm{th}}$ step,
	\begin{enumerate}
		\item For every entangler $\hat{E_i}\in E$, prepare $\hat{E}_i(\vec{\varphi})\ket{\Psi_{n-1}}$ and optimize $\vec{\varphi}$ to obtain the minimum energy estimation. 
		Set the entangler with lowest minimum energy estimation to be $\hat{U}_n$. %
		\item Optimize $\vec{\varphi}_1,\vec{\varphi}_2,\dots,\vec{\varphi}_n$ all together to find the minimum energy estimation for the ansatz $\ket{\Psi_n}= \prod_{j=1}^n \hat{U}_j(\vec{\varphi}_j) \ket{\Psi_0}$. Record the energy estimation.
		\item Go to $(n+1)^{\textrm{th}}$ step and repeat.
	\end{enumerate}
	\item Stop the algorithm based on certain convergence criteria.
\end{enumerate}
Compared with the fixed ansatz methods, by scoring and adaptively adding new entanglers in this way, adaptive methods avoid including irrelevant operations that make small difference to the energy estimation and have the potential to significantly reduce gate counts. 
However, one possible issue of the adaptive method is the size of the entangler pool. 
For instance, the QCC entangler pool consists of the unitaries of the form $\{e^{-i\hat{P}\tau}\}$, where $\hat{P}$ go over all the Pauli words and $\tau$ is an adjustable parameter. 
As the number of Pauli words grow exponentially with the number of qubits they act on, the size of the QCC entangler pool also grows exponentially. 
Although in a recent work~\cite{ryabinkin2020iterative} it was proposed that the ranking of QCC entanglers can be accelerated by partitioning the Hamiltonian into equivalent classes, this problem remains challenging.
Therefore, reduction of the number of entanglers in the pool while preserving numerical accuracy becomes an important pursuit in the adaptive ansatz construction methods. 
\begin{figure}[t]
\centering
	\includegraphics[scale=0.7]{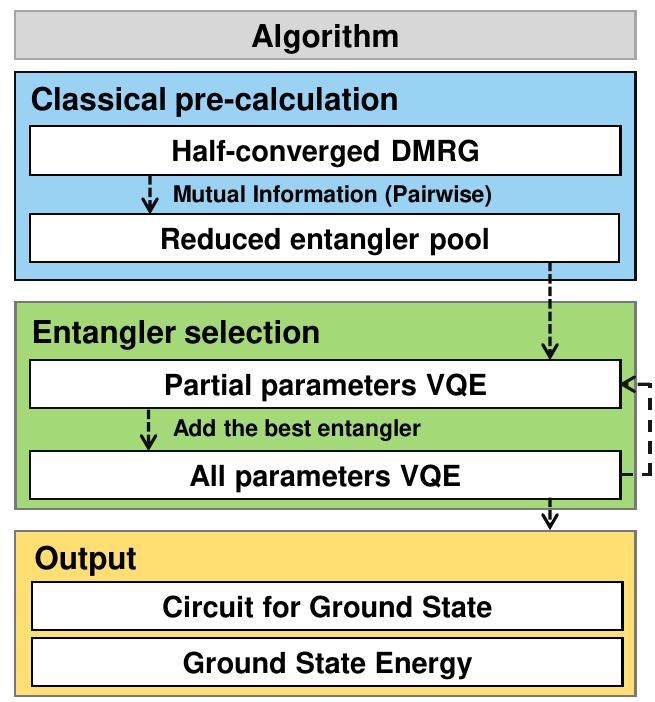}
	\caption{Flowchart of our proposed method. The procedure starts from calculating the mutual information (MI) of the qubits in the approximated ground state obtained from a classical computational method, followed by the adaptive ansatz construction using the entangler pool reduced by MI.}
\label{fig:muqcc_process}
\end{figure}

\subsection{Mutual information-assisted screening}
\label{subsec:MI_screening}
We present our method to reduce the entangler pool size for adaptive ansatz construction here. 
The rationale of our pool reduction protocol is to assign a score to every entangler in the pool based on mutual information (MI)~\cite{amico2008entanglement} and eliminate the entanglers with low scores. The MI between qubit $i$ and $j$ is defined as
\begin{equation}
 \label{equ:MI}
 I_{ij} = \frac{1}{2}\left(S(\hat{\rho}_i)+S(\hat{\rho}_j)-S(\hat{\rho}_{ij})\right),
 \end{equation}
 where $\hat{\rho}_i$, $\hat{\rho}_j$ and $\hat{\rho}_{ij}$ are the reduced density matrices of qubit $i$, qubit $j$ and qubit $i,j$ together, after tracing out the rest of the qubits. 
 $S(\hat{\rho})$ is the von Neumann entropy of a density matrix $\hat{\rho}$ and is defined as
 \begin{equation}
 S(\hat{\rho})=- \sum_i p_i \log_2(p_i),
 \end{equation}
where $p_i$ is the $i^{\textrm{th}}$ eigenvalue of $\hat{\rho}$. 
Here, we propose to use the \textit{correlation strength} as the score of the entanglers.
The correlation strength $C(\hat{E_i})$ of an entangler $\hat{E_i}$ is defined as
\begin{equation}
	C(\hat{E_i})=\frac{1}{L(\hat{E_i})(L(\hat{E_i})-1)}\sum_{j, k\in Q(\hat{E_i});j\neq k} I_{jk},\label{eq:correlation_strength}
\end{equation}
where $Q(\hat{E_i})$ denotes the subset of qubits that $\hat{E_i}$ acts on and $L(\hat{E_i})$ denotes the number of qubits present in $Q(\hat{E_i})$. 
The correlation strength can be regarded as the average MI of qubit pairs within $Q(\hat{E_i})$.
In our method, we first approximate the ground state wavefunction and the MI of all qubit pairs in it by DMRG (or other classical methods). 
Then, we calculate the correlation strength defined in \equref{eq:correlation_strength} for the entanglers and rank the entanglers according to their correlation strengths from high to low.
Finally, we empirically choose first $\pcut$ percent of entanglers and place them into a new reduced pool.
We formally summarize our MI-assisted adaptive VQE above, accompanied by an illustration in Fig.~\ref{fig:muqcc_process}.
\begin{algorithm}[t]
\begin{enumerate}
    \item Prepare the qubit Hamiltonian $\hat{H}$ to solve.
	Set the correlation strength cutoff percentile $\pcut$.
	\item Define the original entangler pool $E_{\text{original}}=\{\hat{E_i}\}$.
	\item Calculate ground state of $\hat{H}$ by DMRG and use the DMRG wavefunction to calculate the MI defined in \equref{equ:MI} of all qubit pairs.
	\item Calculate the correlation strength defined in \equref{eq:correlation_strength} of all the entanglers in the pool. 
	Select $\pcut$ percent of the entanglers that have the largest correlation strength.
	\item Discard all the entanglers in $E_{\text{original}}$ which are not selected in the previous step. Denote the new entangler pool by $E_{\text{screened}}$.
	\item Carry out the adaptive ansatz construction with $E_{\text{screened}}$
\end{enumerate}
\caption{MI-assisted adaptive VQE}
\label{alg:the_algorithm}
\end{algorithm}

\section{Numerical Examples}
\label{sec:results}
\begin{figure*}[ht]
\subfigure[]{
\begin{minipage}[t]{0.48\linewidth}
\centering
\includegraphics[width=0.9\textwidth]{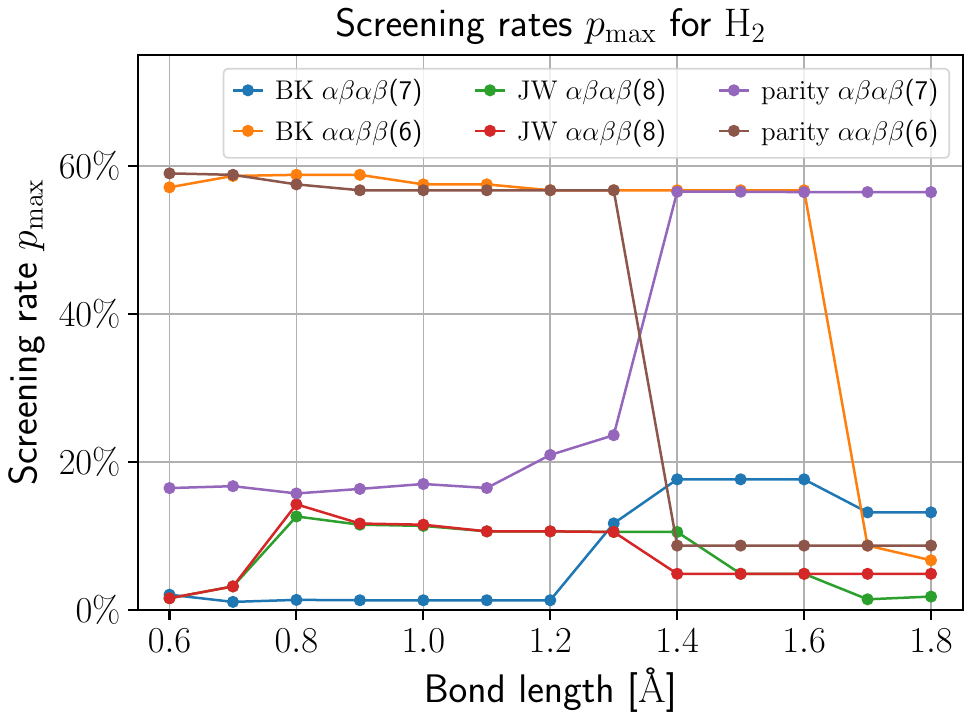}
\end{minipage}
}
\subfigure[]{
\begin{minipage}[t]{0.48\linewidth}
\centering
\includegraphics[width=0.9\textwidth]{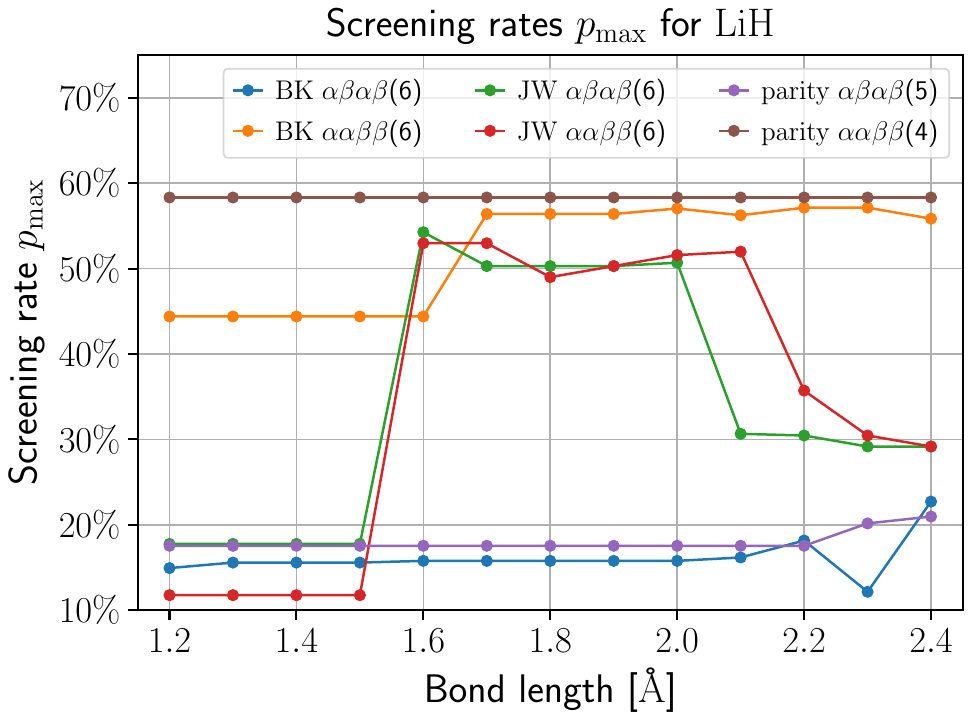}
\end{minipage}
}
\caption{ The screening rate $\pmax$ of the (a) $\mathrm{H_2}$ and (b) $\mathrm{LiH}$ molecule plotted against the bond lengths of the molecules. In our method, by using a screened pool with $\pmax N(E_{\mathrm{original}})$ entanglers, the number of circuit runs needed is reduced to approximately $\pmax$ percent of that in the methods using a complete pool. The number of qubits used in each setting is specified in the parentheses. 
}
\label{fig:max-per}
\end{figure*}

\begin{table*}
\center
\begin{tabular*}{\textwidth}{l@{\extracolsep{\fill}}ccccc}
\hline
molecule                    & $\mathrm{H_2}$  & $\mathrm{LiH}$& $\mathrm{H_2O}$ & $\mathrm{BeH_2}$\\
\hline
molecular configuration     & $d(\mathrm{HH})\in[0.6,1.8]$    & $d(\mathrm{LiH})\in[1.2,2.4]$ & $ d(\mathrm{OH})\in[1.2,2.4] ,\angle \mathrm{HOH}=107.6^{\circ}$ &  $d(\mathrm{BeH})=2.5$ \\
atomic basis set            & 6-31g    & sto-3g   & 6-31g & MRA-PNOs~\cite{doi:10.1063/1.5141880} \\
complete active space (CAS) & 2e/4orb(Full)   & 2e/3orb(A1:3) & 4e/5orb(B2:2,A1:3) & 4e/6orb \\
fermion-to-qubit mapping    & BK,JW,parity   & BK,JW,parity  & parity    & BK \\
number of qubits            & 6,7,8   & 4,5,6   & 8 &11  \\
spin-orbital grouping 		& $\abab, \aabb$  & $\abab, \aabb$ & $\aabb$  & $\abab$  \\
\hline
\end{tabular*}
\caption{Detailed information of the simulated systems. The unit of bond lengths is $\mathrm{\AA}.$\label{tab:calculations}}
\end{table*}
We demonstrate our MI-assisted adaptive VQE on four different molecular systems: $\mathrm{H_2}$, $\mathrm{LiH}$, $\mathrm{H_2O}$ and $\mathrm{BeH_2}$. On every system, we try our method with the QCC entangler pool, in which the entanglers are all the exponentiated Pauli words $\exp(-i\hat{P}\tau)$. Besides, we also try the qubit fermionic-excitation pool from the qubit-ADAPT-VQE method~\cite{tang2019qubit} on the runs of $\mathrm{BeH_2}$, where the system is relatively large.
Because digital quantum computer cannot directly process creation and annihilation operators that are contained in molecular Hamiltonians, a transformation from fermionic Hamiltonians into qubit Hamiltonians is needed.
In this work, Jordan-Wigner~(JW)~\cite{wigner1928paulische}, parity and Bravyi-Kitaev~(BK)~\cite{bravyi2002fermionic} transformations are used and compared. 
Furthermore, because there is no external magnetic field in the systems, we screen all the entanglers containing even number of Pauli $Y$ operators because they commute with $\hat{H}$ \cite{ryabinkin2020iterative}. Details about the simulated systems and the used qubit encodings are given in Table \ref{tab:calculations}. The fermion-qubit transformations are implemented by OpenFermion~\cite{mcclean2020openfermion}; the quantum circuit simulation is implemented by ProjectQ~\cite{steiger2018projectq} and the DMRG calculation is carried out by iTensor~\cite{ITensor}. Parameter optimizations are done by the basin-hopping method~\cite{wales2003energy} from SciPy~\cite{2020SciPy-NMeth} with heuristic parameters of 10 iterations, 0.5 temperature and $10^{-6}$ step size. Chemical accuracy is defined to be 1 milihartree compared with the full configuration interaction (FCI) energy within 
certain basis set and active space. The FCI calculations are done by PySCF~\cite{sun2018pyscf,10.1002/jcc.23981}. 
\begin{figure*}[t]
\subfigure[]{
\begin{minipage}[t]{0.48\linewidth}
\centering
\includegraphics[width=0.9\textwidth]{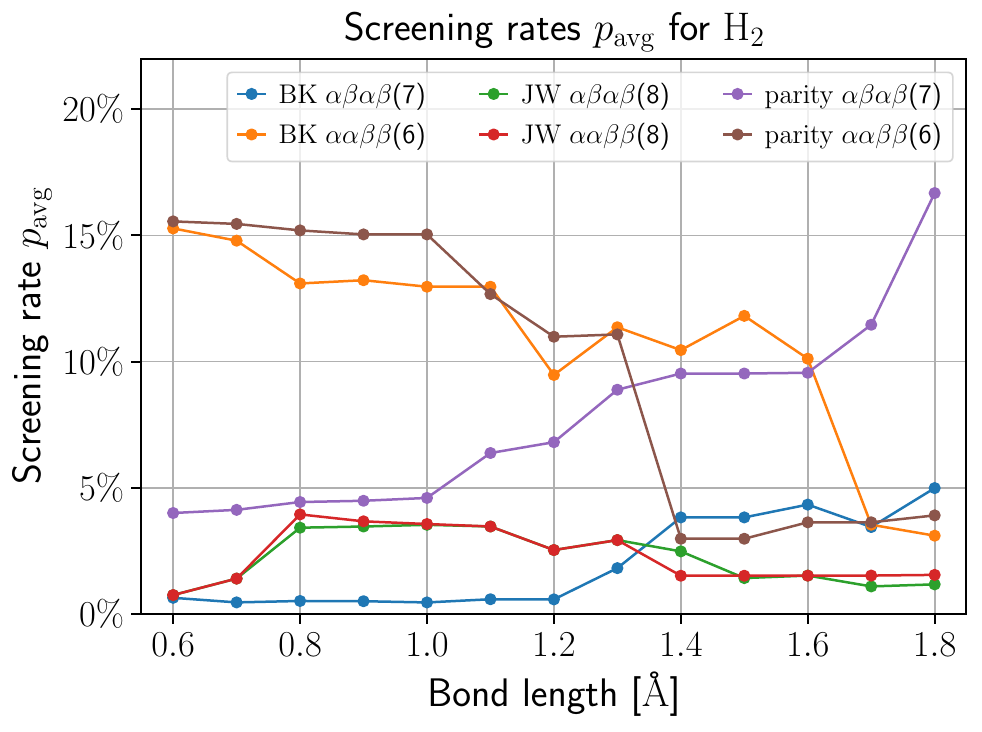}
\end{minipage}
}
\subfigure[]{
\begin{minipage}[t]{0.48\linewidth}
\centering
\includegraphics[width=0.895\textwidth]{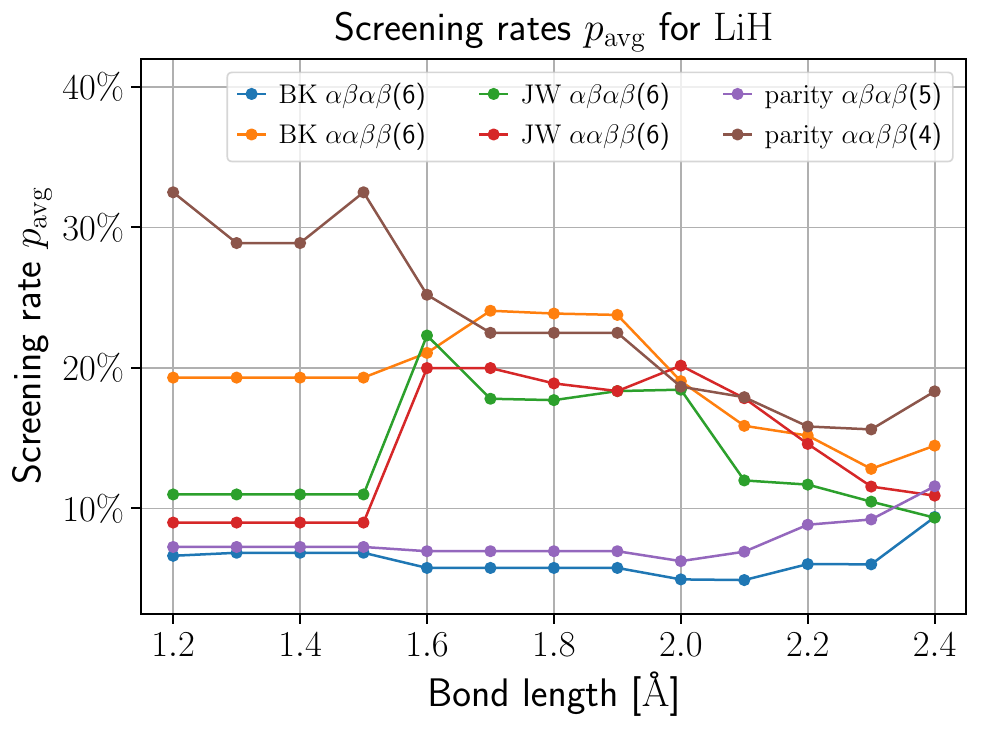}
\end{minipage}
}
\caption{The screening rate $\pavg$ of the (a) $\mathrm{H_2}$ and (b) $\mathrm{LiH}$ molecule plotted against the bond lengths of the molecules. The low $\pavg$ of a run implies that the entanglers added in the circuit of the run tend to have high correlation strength overall. The number of qubits used in each setting is specified in the parentheses.
}
\label{fig:avg-per}
\end{figure*}

To evaluate the performance of our method, we need to estimate how the correlation strength of an entangler is related to its possibility to be added in the adaptive ansatz construction process.
If the correlation strength of the selected entanglers are relatively high, removing the entanglers with low correlation strength would make no difference to the ansatz construction because they will not be added in the ansatz.
For this reason, we define a quantity, \textit{correlation percentile}, or simply percentile, for entanglers present in an entangler pool.
For an entangler $\hat{E}_i$, suppose there are $N_{\geq}(\hat{E_i})$ entanglers whose correlation strength is larger or equal to $\hat{E_i}$'s. 
We define its percentile $\per (\hat{E_i})$ as $N_{\geq}(\hat{E_i})/N(E_{\mathrm{origin}})$, where $N(E)$ denotes the size of the entangler pool $E$. In our numerical experiments, we record the percentile of every selected entangler and present analysis of a simulation run based on them.
Especially, we define and track the screening rates $\pmax$ and $\pavg$. They are computed by the percentiles of the entanglers in the final circuit after convergence in a run is reached.
Suppose $\{\hat{U_i}\}$ are the entanglers added to the ansatz when the adaptive construction stops. We define $\pmax$ as
\begin{equation}
    \pmax=\max_i \per(\hat{U_i}),
\end{equation}  
which is the highest percentile of the entanglers in the ansatz. $\pmax$ can also be understood as the percentile of the selected entangler with lowest correlation strength in the final circuit. Similar to $\pmax$, we also define $\pavg$ as 
\begin{equation}
    \pavg=\frac{1}{N_{\mathrm{ent}}} \sum_i \per(\hat{U_i}),
\end{equation}
which is the average value of the percentiles of all the selected entanglers $N_{\mathrm{ent}}$. Both $\pmax$ and $\pavg$ represent how the correlation strength of the selected entanglers are larger than that of the unselected ones, and provide means to quantify speed-ups compared with methods using a complete entangler pool. Therefore, we call them screening rates. 
As long as $\pcut$ is set to be higher than $\pmax$, the entangler selection at each step will be the same as using the original pool. 
A low $\pmax$ means a low $\pcut$ can be set and one only needs to try a small portion of entanglers in the original pool. 
Suppose $N_{\mathrm{ent}}$ entanglers were to be added eventually. 
With the knowledge of $\pmax$, one would need to carry out $\pmax N(E_{\mathrm{origin}})$ entangler trials at each step.
In total, we only need to carry out $N_{\max}=\pmax N(E_{\mathrm{origin}}) N_{\mathrm{ent}}$ entangler trials in the entire calculation.
As to $\pavg$, unlike $\pmax$, we cannot use a fixed screened pool with $N_{\mathrm{avg}}=\pavg N(E_{\mathrm{original}})$ entanglers to recover the entangler selection using the original pool because in some steps we need a pool with $N_{\max}$ entanglers to recover the selection. However, we expect that by adjusting the screened pool at each step using some sophisticated techniques, with an average pool size between $N_{\mathrm{avg}}$ and $N_{\max}$, one can recover the original entangler selection.

\begin{figure*}[t]
\subfigure[]{
\begin{minipage}[t]{0.48\linewidth}
\centering
\includegraphics[width=0.88\textwidth]{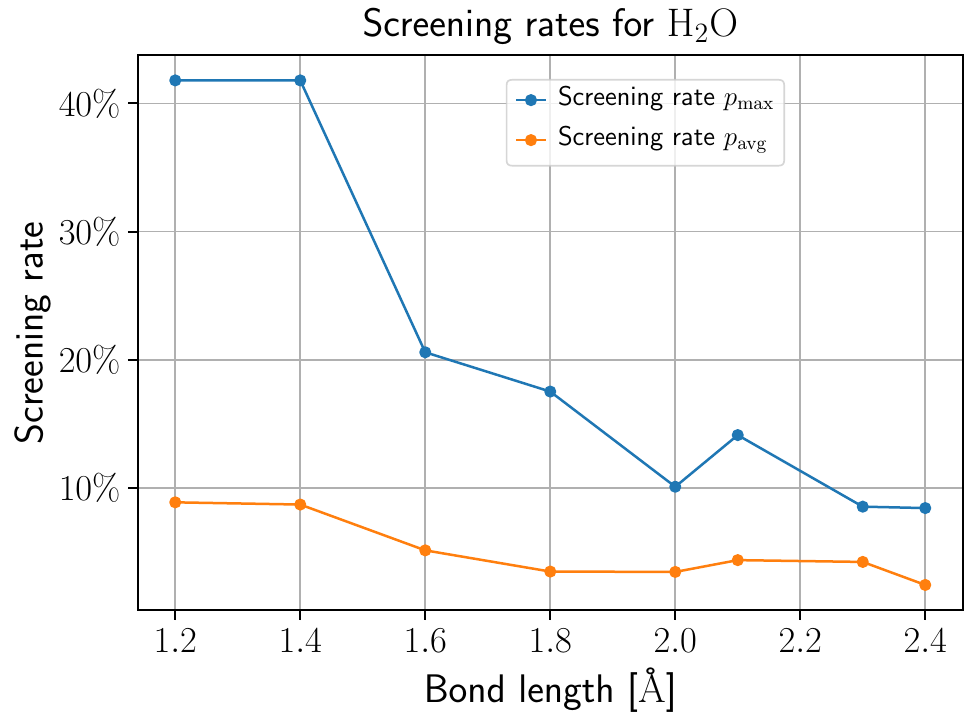}
\end{minipage}
}
\subfigure[]{
\begin{minipage}[t]{0.48\linewidth}
\centering
\includegraphics[width=0.9\textwidth]{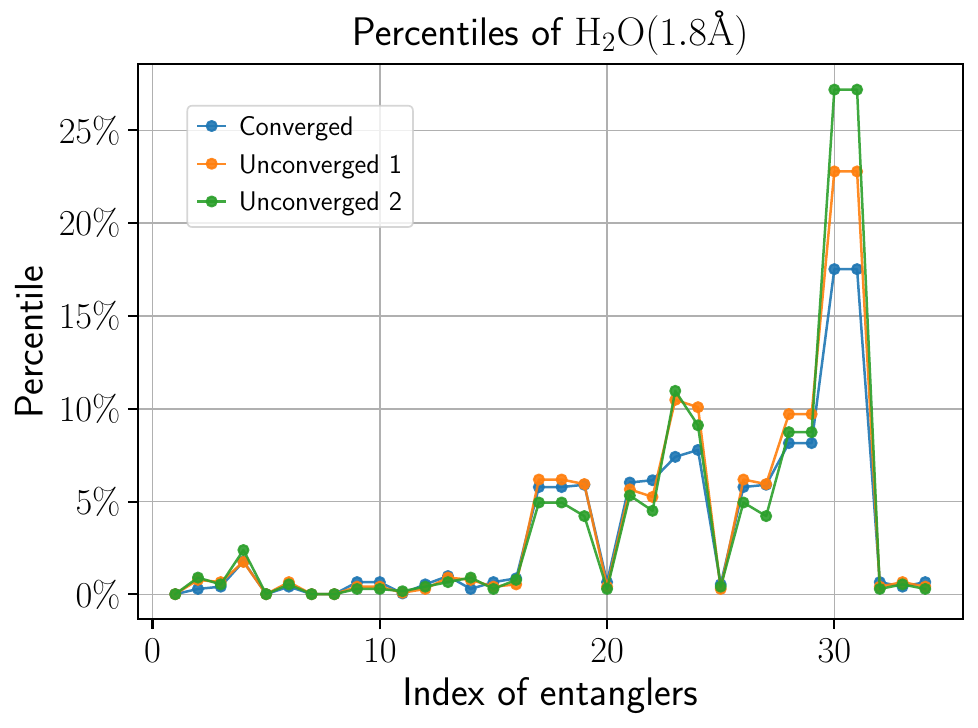}
\end{minipage}
}
\caption{(a)~The screening rate $\pmax$ and $\pavg$ for the $\mathrm{H_2O}$ molecule with different bond lengths.  \label{fig:h2o-per}
(b)~Percentiles of entanglers in the run for $\mathrm{H_2O}$ with bond length of $1.8$ $\AA$. The index represents the order of adding the entanglers. The entangler with smaller index is added earlier in the run.
	The energy of the run \textit{Unconverged 1} and \textit{Unconverged 2} are about 10 milihartree and 20 milihartree away from the (Converged) FCI energy respectively.
	\label{fig:h2o-pauli_per}
}
\end{figure*}

We first run the adaptive construction with complete pools ($\pcut=100\%$) to show the percentiles of the selected entanglers and how low $\pcut$ can be set for the systems of $\mathrm{H_2}, \mathrm{LiH}$ and $\mathrm{H_2O}$. 
As we are only interested in whether the entanglers of low percentile are likely to provide a relatively large energy descents and do not concern whether their energy descents are the largest, in these runs, we do not insist on adding the entangler whose energy descent is the largest at each step.
Instead, we regard all the entanglers that give an energy descent larger than $30\%$ of the largest energy descent as acceptable entanglers and then add the entanglers with the largest correlation strength among all the acceptable entanglers. In the runs of $\mathrm{BeH_2}$, as we are interested in how percentile cutoff affects the selection of entanglers, this strategy is no longer adopted. In all of our numerical experiments, we terminate the calculations when the chemical accuracy is reached.

\subsection{$\mathrm{H_2}$ and $\mathrm{LiH}$ molecules}
To begin with, we carry out our method with the QCC pool on $\mathrm{H_2}$ and $\mathrm{LiH}$ with basis and active space specified in Table \ref{tab:calculations}. 
The bond lengths are chosen from 0.6 $\AA$ to 1.8 $\AA$ for $\mathrm{H_2}$ and 1.2 $\AA$ to 2.4 $\AA$ for $\mathrm{LiH}$, respectively. 
The orbitals are ordered by their energy before any fermion-qubit transformation.
We compare the results obtained from different transformations as well as different grouping of spin-orbitals before the transformation. 
Particularly, we study the $\aabb$ and $\abab$ groupings. 
In the $\aabb$ grouping, spin-orbitals of $\alpha$ spin are placed first, followed by the $\beta$ spin.
In the $\abab$ grouping, spin-orbitals with $\alpha$ and $\beta$ spins are placed in alternating order. 
In the runs, the number of qubits needed depends on the fermion-qubit transformation and the size of the entangler pools of the runs are determined by its qubit number. The run with 4, 5, 6, 7 and 8 qubits uses a pool with 120, 496, 2016, 8128 and 32640 entanglers, respectively.

\begin{figure*}[t]
\subfigure[]{
\begin{minipage}[t]{0.48\linewidth}
\centering
\includegraphics[width=0.88\textwidth]{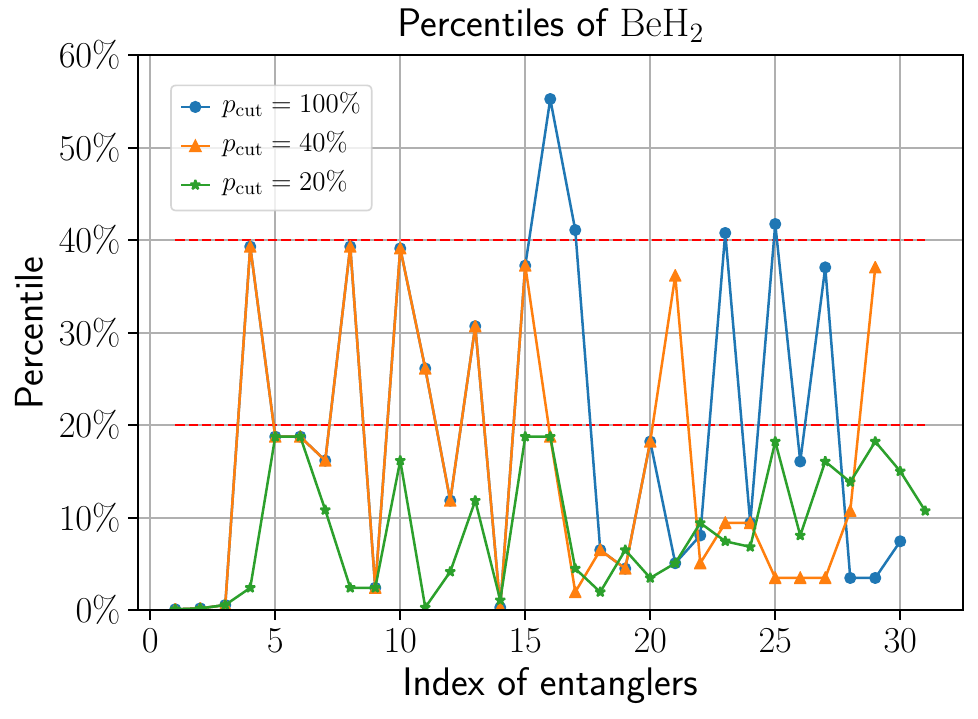}
\end{minipage}
}
\subfigure[]{
\begin{minipage}[t]{0.48\linewidth}
\centering
\includegraphics[width=0.9\textwidth]{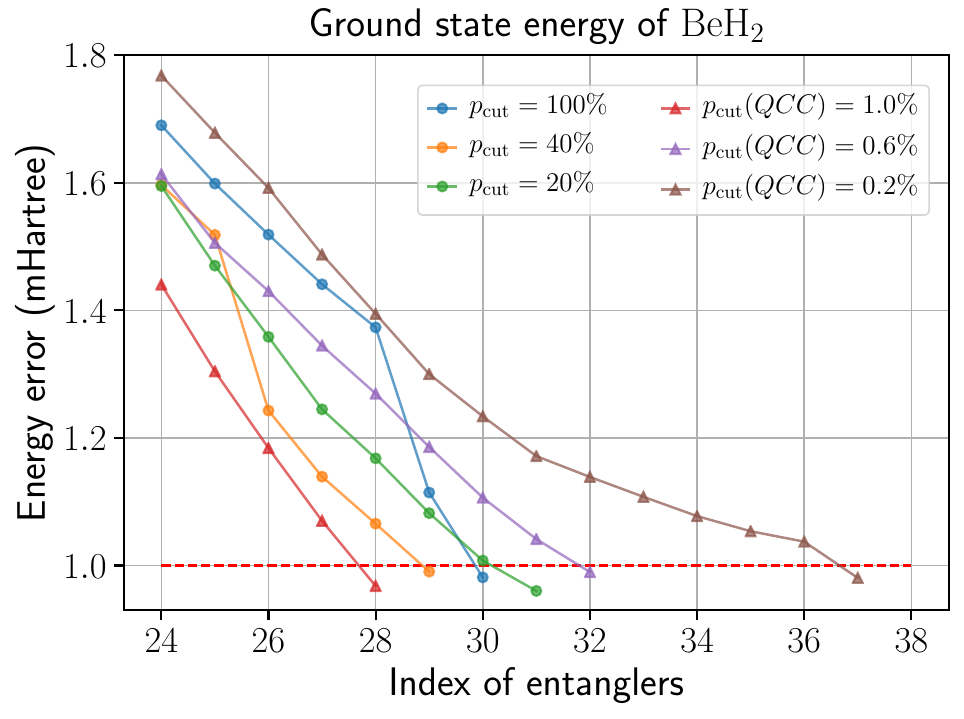}
\end{minipage}
}
\caption{(a) The percentiles of the selected entanglers in the runs of $\beh$ with $\pcut=20\%,40\%,100\%$ on the qubit fermionic-excitation pool. Some entanglers with larger energy descent are excluded because of the cutoffs. However, the circuit depths for chemical accuracy are not increased significantly in this case. (b) The errors of the ground state energies in each iterations of the runs with both the qubit fermionic-excitation pool and QCC pool. The not specified runs are with qubit fermionic-excitation pools and the red line represents the chemical accuracy. The Hamiltonian for the simulation is generated by the MRA-PNO-MP2 method~\cite{doi:10.1063/1.5141880} for compact representation of large quantum systems. We find that the screening increases the depth of circuits in the runs with QCC pools.
\label{fig:beh2_per}
}
\end{figure*}

Our numerical simulation results are presented in Figs.~\ref{fig:max-per} and \ref{fig:avg-per}. 
We find that the resulting percentiles can be very low for both $\mathrm{H_2}$ and $\mathrm{LiH}$ molecules in some cases, meaning we can neglect most parts of the original pool and a significant speed-up can be achieved. 
For example, in the case of $\mathrm{H_2}$ with BK $\abab$, $\pmax$ is around $1.5\%$ for some bond lengths (see Fig.~\ref{fig:max-per}(a)), showing that in these settings high correlation strength strongly correlates with high energy descent. 
It also implies that the original entangler pool can be reduced to $1.5\%$ of its original size.
Besides $\pmax$, the $\pavg$ of the runs, which are less than half of the corresponding $\pmax$ in most of the cases, also show that the entanglers in these runs are mainly placed among the qubits with high correlation strength. 
We notice that our method gives disparate screening rates for different transformations and bond lengths, and in some cases the screening rate $\pmax$ is higher than $50\%$.
However, we like to emphasize that, for most of the systems with high screening rate, the pools have already been reduced by removing stationary qubits and the screening rates are calculated with respect to the already reduced pools.

If the screening rates were computed with respect to the original 8-qubit ($\mathrm{H_2}$) and 6-qubit ($\mathrm{LiH}$) pools, BK $\abab$ still presents lowest $\pmax$ for $\mathrm{H_2}$ in bond length less than or equal to $1.3$ $\AA$. But for $\mathrm{LiH}$ and $\mathrm{H_2}$ in bond length larger than $1.3$ $\AA$, the encoding with lowest $\pmax$ becomes parity $\aabb$. The screening rates computed in this way are shown in the Appendix \ref{Appendix}.
A detail worth mentioning here is that the orbitals of $\mathrm{H_2}$ have different irreducible representations and a orbital ordering change happens in the bond length interval from $1.3$ $\AA$ to $1.4$ $\AA$. This might be in part responsible for the change of $\pmax$ there for some transformations.
In contrast, the orbitals of $\mathrm{LiH}$ all have the same irreducible representations. 
Based on the above observation, we believe that a system-dependent proper transformation, concerning spin-orbital grouping and orbital ordering, should be chosen for an improved implementation of our algorithm. 
For a clearer picture, more in-depth benchmarks and studies are required.

\subsection{$\mathrm{H_2O}$ molecule}

We further evaluate the performance of our method on $\mathrm{H_2O}$ with bond angle $\angle\mathrm{HOH}=107.6^{\circ}$ and bond length ranging from 1.2 $\AA$ to 2.4 $\AA$. The basis and transformation used is 6-31g and the parity transformation. 8 qubits are needed after choosing an active space specified in Table \ref{tab:calculations} and reducing 2 stationary qubits by $\aabb$ grouping combined with the parity transformation~\cite{bravyi2017tapering}. The number of entanglers in the complete 8-qubit QCC pool is $32640$. We add a spin penalty term $S^2$ to the original Hamiltonian to help the calculation converge when the bond length is larger or equal to $1.8$ $\AA$. In this setting, $\pmax$ and $\pavg$ both show a decreasing trend as the bond length grows. Especially, $\pmax$ and $\pavg$ for bond length $2.4$ $\AA$ is only $8.44\%$ and $2.44\%$ (see Fig.~\ref{fig:h2o-per}(a)), meaning that our method can largely reduce the size of the entangler pool for the stretched water system, where the correlation energy is large.

Due to the relatively large correlation energy present in $\mathrm{H_2O}$, the numbers of entanglers needed to converge to chemical accuracy are larger than that of $\mathrm{H_2}$ and $\mathrm{LiH}$, providing good examples to show the stability of percentiles in our method with inaccurate MI. 
Let us take the water molecule with bond length 1.8 $\AA$ as an example. In this case, 34 entanglers are used for chemical accuracy. In Fig.~\ref{fig:h2o-pauli_per}(b), we show how the percentile varies by using MI from DMRG calculations of different convergence level. 
We included three sets of results, one for a converged calculation whose energy is within 1 milihartree from the FCI energy, one about 10 milihartree and one about 20 milihartree from the FCI energy. We find that for the first 16 entanglers, where the percentiles are comparatively low, the differences among all the three results are not significant. 
The deviation becomes apparent when the percentile is high. 
In the worst case, the MI from the unconverged DMRG calculation only lift the $\pmax$ of this run from $17.5\%$ to $27.2\%$. Considering the large deviation of energy from the FCI, we claim that in this run unconverged DMRG still provides a good MI estimation. 
This is vital for our algorithm to be applied to larger physical systems where accurate estimations of MIs are expected to be challenging.

\subsection{$\beh$ molecule}

To test the performance of our screening protocol on a larger system and smaller entangler pools, we simulate our method on the $\beh$ molecule with bond length $2.5\mathrm{\AA}$.
In this example, we generate the qubit Hamiltonian by a surrogated basis-set-free approach called MRA-PNO-MP2~\cite{doi:10.1063/1.5141880} according to Ref.~\cite{kottmann2020reducing} using \textsc{tequila}~\cite{kottmann2020tequila} with \textsc{madness}~\cite{harrison2016madness} as backend.
The active orbitals that define the qubit Hamiltonian are two occupied Hartree-Fock orbitals and four optimized pair-natural orbitals on MP2 level. After tapering one stationary qubit by the BK transformation, the qubit Hamiltonian of the system contains 11 qubits. 
Adaptive ansatz construction on Hamiltonian generated in this way have been studied in a previous work~\cite{kottmann2020reducing} and here we apply our method with both the QCC pool and the qubit fermionic-excitation pool on it.
The latter was first proposed in the qubit-ADAPT-VQE method~\cite{tang2019qubit} and here we define it as the pool containing all the entanglers $\{e^{-i\hat{P}_it}\}$ whose Pauli words $\{\hat{P}_i\}$ appear in the transformed set of fermionic operators $\{a_q^\dagger a_p^\dagger a_ra_s-a_s^\dagger a_r^\dagger a_pa_q\}\cup\{a_q^\dagger a_p-a_p^\dagger a_q\}$  with $q,p,r,s$ going over the indices of all the spin-orbitals. 
We discard all the entanglers whose Pauli word contains the stationary qubit. In total, there are $4137$ entanglers in the qubit fermionic-excitation pool, which is much less than the QCC pool containing $2.10\times 10^6$ entanglers. 
To show how the percentile cutoff affects the selection of entanglers, we screen the qubit fermionic-excitation pool with $\pcut=20\%,40\%$ and $100\%$~(no cutoff) and screen the QCC pool with $\pcut=0.2\%,0.6\%$ and $1.0\%$. The used pools contains $1002$, $1638$, $4137$ for the qubit fermionic-excitation pool and $4102$, $12544$, $20906$ entanglers for the QCC pool respectively. We show the result of the runs in Fig.~\ref{fig:beh2_per}. The percentiles for the runs with the qubit fermionic-excitation pool are shown in Fig.~\ref{fig:beh2_per}(a).
We find that the percentile cutoff does exclude some entanglers with larger energy descent from being added. However, as shown in Fig.~\ref{fig:beh2_per}(b), selecting entanglers with smaller energy descent does not significantly affect the circuit depth required for chemical accuracy for the qubit fermionic-excitation pool. The required numbers of entanglers in the circuits are $31,29$ and $30$ for $\pcut=20\%,40\%$ and $100\%$, which implies the entanglers in the larger pool may make the circuits for convergence even deeper. For the runs on the QCC pool, we find the numbers of entanglers for convergence is $37,32$ and $28$ for $\pcut=0.2\%,0.6\%$ and $1.0\%$, showing that the final circuit depths are reduced by using larger QCC pools. We conjecture this is because the QCC pools contains more critical entanglers which can reduce the circuit depth for convergence but are not in the qubit fermionic-excitation pool. 
We also like to remark that our simulations produce much shallower quantum circuits compared with the previous work~\cite{kottmann2020reducing} using a UCC pool. Though the UCC pool is much smaller and contains only $345$ entanglers, $31$ UCC entanglers, which are equivalent to $200$ QCC entanglers, are needed in the converged circuit in this case. 
By screening the tested pools, our MI-assisted adaptive VQE mitigates the problem of large pool size for them and thus helps produce compact VQE circuits more efficiently.

\section{Conclusions \& Discussion}
\label{sec:conclusions}

In this work, we provide a method to reduce the size of the entangler pools for adaptive ansatz construction methods. The new method uses mutual information which can be approximated by existing classical algorithms such as DMRG. Our method reveals how MI can help to construct the VQE ansatz and allow adaptive ansatz construction with much less circuit runs.
We also find that the average correlation strength $\pavg$ can be very low, which implies that the entanglers should be mainly placed among the qubits in which the MI between the qubits is high. 
Furthermore, the screening rates are not independent of the chosen fermion-qubit transformation and disparate $\pmax$ and $\pavg$ are observed for different transformations, providing an additional criterion to quantify the performance of different transformations~\cite{doi:10.1021/acs.jctc.8b00450}.

A potential drawback of the proposed method is the absence of a systematic way to predict $\pmax$ and set $\pcut$.
Here, we list some strategies that could be used to heuristically set $\pcut$ in practice. 
\begin{enumerate}
    \item $\pcut$ can be chosen based on the required accuracy, circuit depth the hardware can support and the run time one is willing to spend. One can also test the stability of the result under a slight change of $\pcut$.
    \item $\pcut$ can also be decided based on the selection of entanglers in the runs. 
    If many entanglers of very high percentiles (like 90\% of the cutoff) are selected, one may consider choosing a larger $\pcut$.
    \item As we find that $\pmax$ can be nearly continuous with respect to the geometry of the molecular system (see Fig.~\ref{fig:h2o-per}(a)), if suitable $\pcut$ values are known for a range of chemical systems, similar values might be used for systems that are expected to behave similarly.
\end{enumerate}

In addition, although low percentile means high speed-up, it remains unknown that whether low percentile means also low difficulty for classical computers to simulate. 
It will be interesting to bridge the two percentile values $\pmax$ and $\pavg$ we have defined in this work with classical simulation resource count. 
If low $\pmax$ was associated with low difficulty in classical simulation, $\pmax$ could be a new quantifier to indicate the difficulty of simulating a system classically.
If the converse was true, a system with low $\pmax$ and high difficulty to simulate classically would be a good problem choice to demonstrate quantum advantage over the classical counterpart.

Finally, we concur that the pre-calculation of mutual information may provide a valuable recipe for running VQE on noisy quantum computers with limited connectivity and limited capabilities. 
As the pre-calculation with classical computers suggests how much entanglement should be produced among qubit pairs, one may map the qubits to physical noisy qubits according to their connectivity and ability to produce entanglement in the optimal hardware configuration~\cite{tkachenko2020correlation}.

\section*{Acknowledgements}
T.H.K., J.K., M.D. and A.A.-G. acknowledge funding from Dr. Anders G. Fr{\o}seth.
A.A.-G. also acknowledges support from the Canada 150 Research Chairs Program and from Google, Inc. in the form of a Google Focused Award.
M.D. also acknowledges support by the U.S. Department of Energy, Office of Science, Office of Advanced Scientific Computing Research, Quantum Algorithms Teams Program.
This work was supported by the U.S. Department of Energy under Award No. DE-SC0019374. 
The authors declare that there are no competing interests.

\bibliographystyle{apsrev4-1}
\bibliography{MI_Adapt_VQE}

\clearpage

\appendix
\onecolumngrid
\section{}
\label{Appendix}

\begin{figure*}[h]
\subfigure[]{
\begin{minipage}[t]{0.48\linewidth}
\centering
\includegraphics[width=0.9\textwidth]{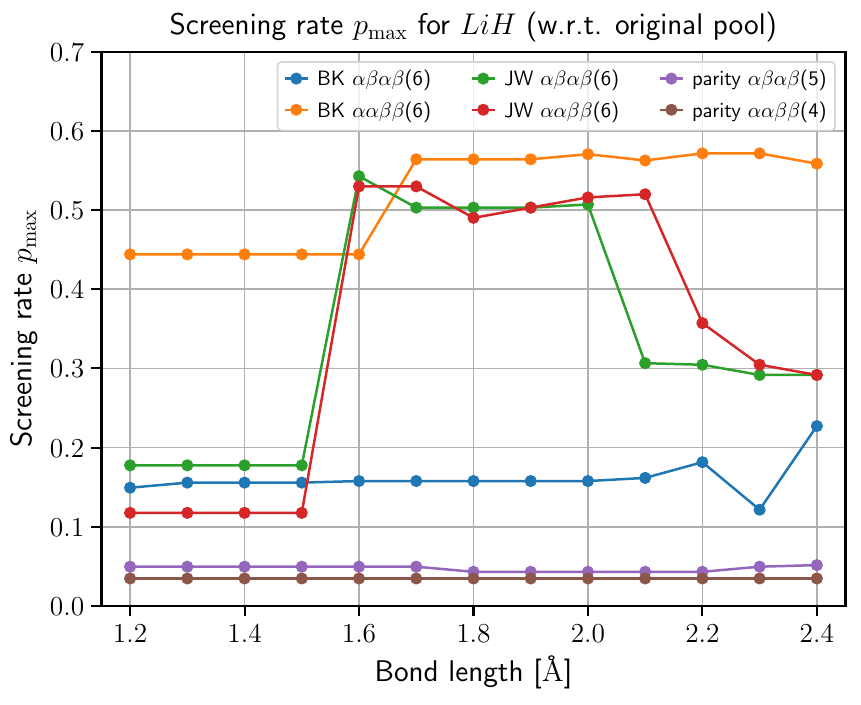}
\end{minipage}
}
\subfigure[]{
\begin{minipage}[t]{0.48\linewidth}
\centering
\includegraphics[width=0.9\textwidth]{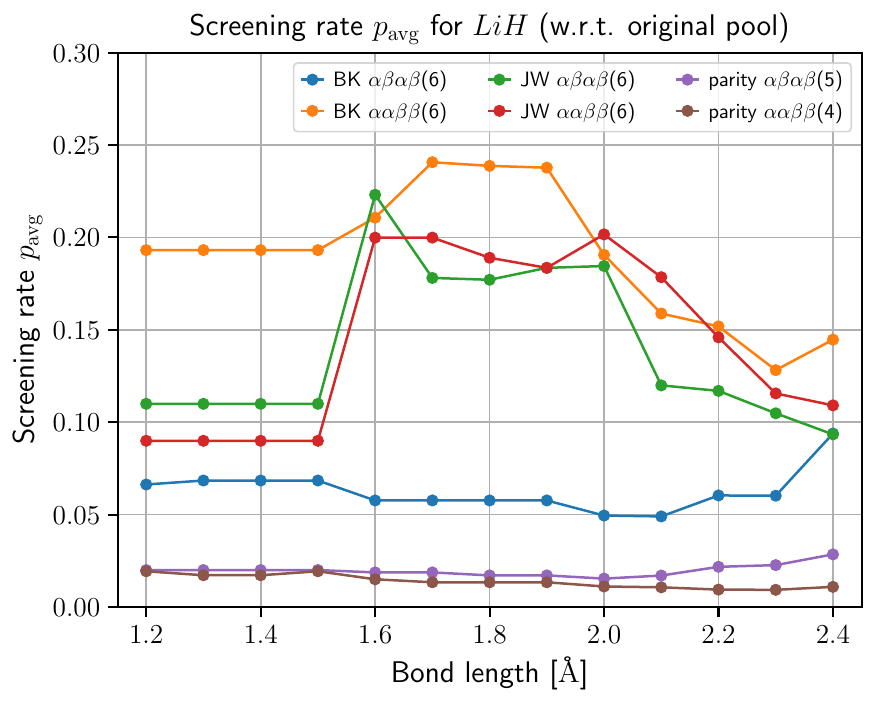}
\end{minipage}
}
\caption{The screening rates $\pmax$ and $\pavg$ for $\mathrm{LiH}$ with respect to the 6-qubit original pool (2016 entanglers).
}
\end{figure*}

\begin{figure*}[h]
\subfigure[]{
\begin{minipage}[t]{0.48\linewidth}
\centering
\includegraphics[width=0.9\textwidth]{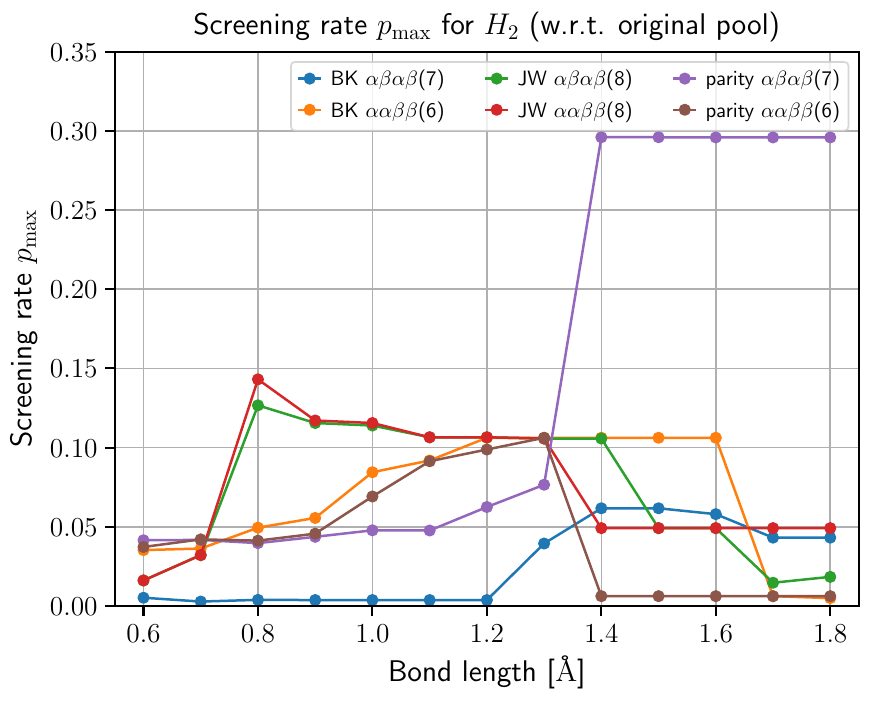}
\end{minipage}
}
\subfigure[]{
\begin{minipage}[t]{0.48\linewidth}
\centering
\includegraphics[width=0.9\textwidth]{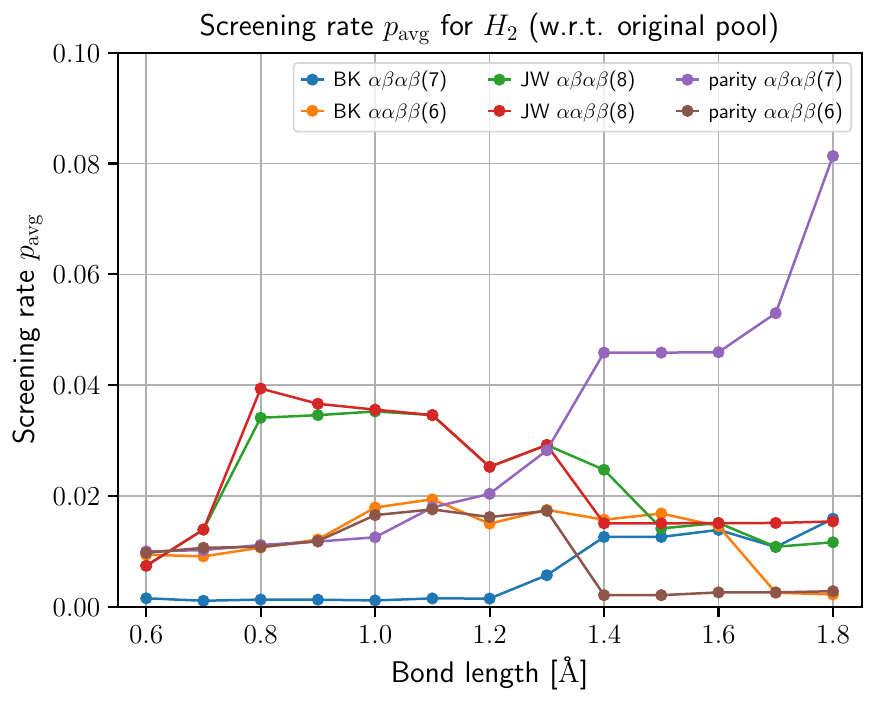}
\end{minipage}
}
\caption{The screening rates $\pmax$ and $\pavg$ for $H_2 $with respect to the 8-qubit original pool (32640 entanglers).
}
\end{figure*}

\end{document}